# Chapter 6

## On the Superluminal Quantum Tunneling and "Causality Violation"

M. Fayngold
*Department of Physics, New Jersey Institute of Technology, Newark, NJ 07102*
faygold@mailaps.org

This is an analysis of some aspects of an old but still controversial topic, superluminal quantum tunneling. Some features of quantum tunneling described in literature, such as definition of the tunneling time and a frequency range of a signal, are discussed. The argument is presented that claim of superluminal signaling allegedly observed in frustrated internal reflection experiment was based on the wrong interpretation of the tunneling process. A thought experiment similar to that in the Tolman paradox is discussed. It shows that a new factor, attenuation, comes in the interplay between tunneled signals and macroscopic causality.

### 1. Introduction (statements and citations)

In discussing the superluminal quantum tunneling (SQT) effect, it is necessary to emphasize from start that contrary to a widely-spread misconception, motions faster than light (FTL) are totally consistent with Special Relativity (SR), and are well described within its framework. Some known examples include superluminal phase velocity of electromagnetic waves in plasma and their superluminal group velocity within a resonant absorption band. The basic feature of all such motions is that they cannot be harnessed for superluminal signaling (SS). Such signaling is indeed, forbidden because it would contradict causality. The ban on SS avoids this contradiction and is frequently referred to as *relativistic* causality. With SS banned, the SQT effect would easily find its niche in physics. The discourse appeared only as a reaction to claims [1-5] that SQT includes SS as its basic feature. But before considering the arguments for SS presented in [1-5], it is worthwhile to show the way its proponents discuss some opposing conventional views.

A few years ago I came across a paper [1] by G. Nimtz and A. A. Stahlhofen. The authors wrote: "*…recently an argument has been raised that zero time inside barriers and thus a superluminal signal speed can never happen…*". Similar thing was said in book [2] (Preface, XVI) by G. Nimtz and A. Haibel: "*… there are some physicists like Moses Fayngold, who tried hard to show that quantum mechanical tunneling does not violate special relativity*." Both quoted statements refer to my first book [6] on SR; but ironically, precisely the topic of SQT *has not been discussed* in that book, so the authors reacted to what I had not yet written by that time. Also, as stressed in the beginning, SR does *not* prohibit FTL motions as such. This has been clearly stated in [6] and vividly described in [7]. As emphasized in [8] on another occasion, the mathematical structure of SR explicitly admits the possibility of superluminal particles (tachyons [9-15]), and moreover, it is *begging* for their existence to satisfy some specific symmetry requirements [6, 16]. The SS is prohibited only because reaching into the past of a process by means of SS can act to eliminate its future (the Tolman paradox [17]). So the authors of [1, 2] have put into my mouth the statements that I had never made. Nevertheless, I had restrained from any comments at that time.

Recently a colleague [18] turned my attention to an article [3] criticizing the natural scientific skepticism about reports on experimental realization of SS. This article also contains citations from [6] skillfully taken out of context and thus totally distorting the views presented there.

Had there been only one or two misleading references, one could attribute them to coincidental errors which we all occasionally make. But a few separately published inaccurate or false statements on the same topic warrant response.

As already mentioned, the SQT topic is absent in [6]. Moreover, I was even (fairly!) criticized by J. Cramer [19] for the "*conspicuous absence*" (using his own words) of this important topic. My first addressing this phenomenon was yet to be published in another book [16] only in the end of June of 2008. So, the above statements in [1, 2], given the time necessary for their publication, were made at least *a few months before* my first mentioning superluminal tunneling in any public or private correspondence. In this respect, they look like the first evidence of an effect emerging before its cause. In other words, they look like a signature of causality violation. And this happened on macroscopic level – all relevant facts are publicly recorded and easy to check! This groundbreaking event makes all SQT experiments hopelessly outdated.

On a more serious note, the demonstrated misleading statements in [1, 2] sit well together with the inaccurate quotations in [3]. Here is one of them sighting a paragraph from [6]:

"*Because all observable properties of material objects are real*, *the appearance of imaginary values in the theory indicates that corresponding quantities cannot be measured. But what cannot*, *in principle*, *be observed*, *does not exist. In other words, there cannot be any superluminal particles.*"

This is an exact clone from the original. But we all know that even a correct quote taken out of context may acquire the opposite meaning. The discussed case is a good example. To avoid such cases, it is customary, when citing a paragraph from a book, to indicate the page or at least the chapter. But precisely in this case, the authors of [3] forgot to do it. I have no choice but to do it myself here, in order to restore the truth. The quoted paragraph is taken from book [6], p. 228. The curious reader with access to the book can open it on this page and read the cited paragraph. But then read the next one! This will change the meaning of the citation to the opposite.

Now I turn from these technical details just illustrating the level of presentation of some facts in [1-3] to a more important aspect – the actual physical arguments about SQT presented there.

## 2. Some features of SQT

### 2.1 *Tunneling time and scattering*

One of the most fundamental properties of SQT is the time it takes the particle to tunnel through a potential barrier. The concept of tunneling time (TT) has many aspects and is far from trivial [20-23]. There are ongoing debates on this topic. Unfortunately, even in [3], which is one of the latest papers on SQT, there is no clear description of TT. In [1], the authors use two different definitions of TT, without discussing how (if at all!) they are related to each other, so the result is total confusion. Let us first focus on definition they use in the introductory part:

"*…the tunneling time τ…is given by a scattering time at the barrier entrance*".

One cannot dispute a definition, but one can always question its relevance to a discussed phenomenon. In the given case, such relevance is, at best, doubtful. We cannot separate the initial wave traversing the barrier from the elastic forward-scattered wave [24-26], since they are coherent and the measurable characteristic (transmitted wave) is their superposition (see also the optical theorem for planar scattering [27]). In addition, the observable outcome for a rectangular barrier is determined by *multiple* scattering on *both* sides [24, 25, 28]. Consider, for instance, the case most favorable for separating the waves scattered on different sides of the barrier – when the incident wave packet is much narrower than the barrier width *d*,

$$\Delta x \ll d \qquad (1)$$

Then the output may be a succession of the secondary packets on both – the incidence and the transmission sides of the barrier. They may undergo some additional spread due to different phase shifts for different spectral components during reflection. So the wave shapes formed on the front and back sides will partially overlap (Fig. 1).

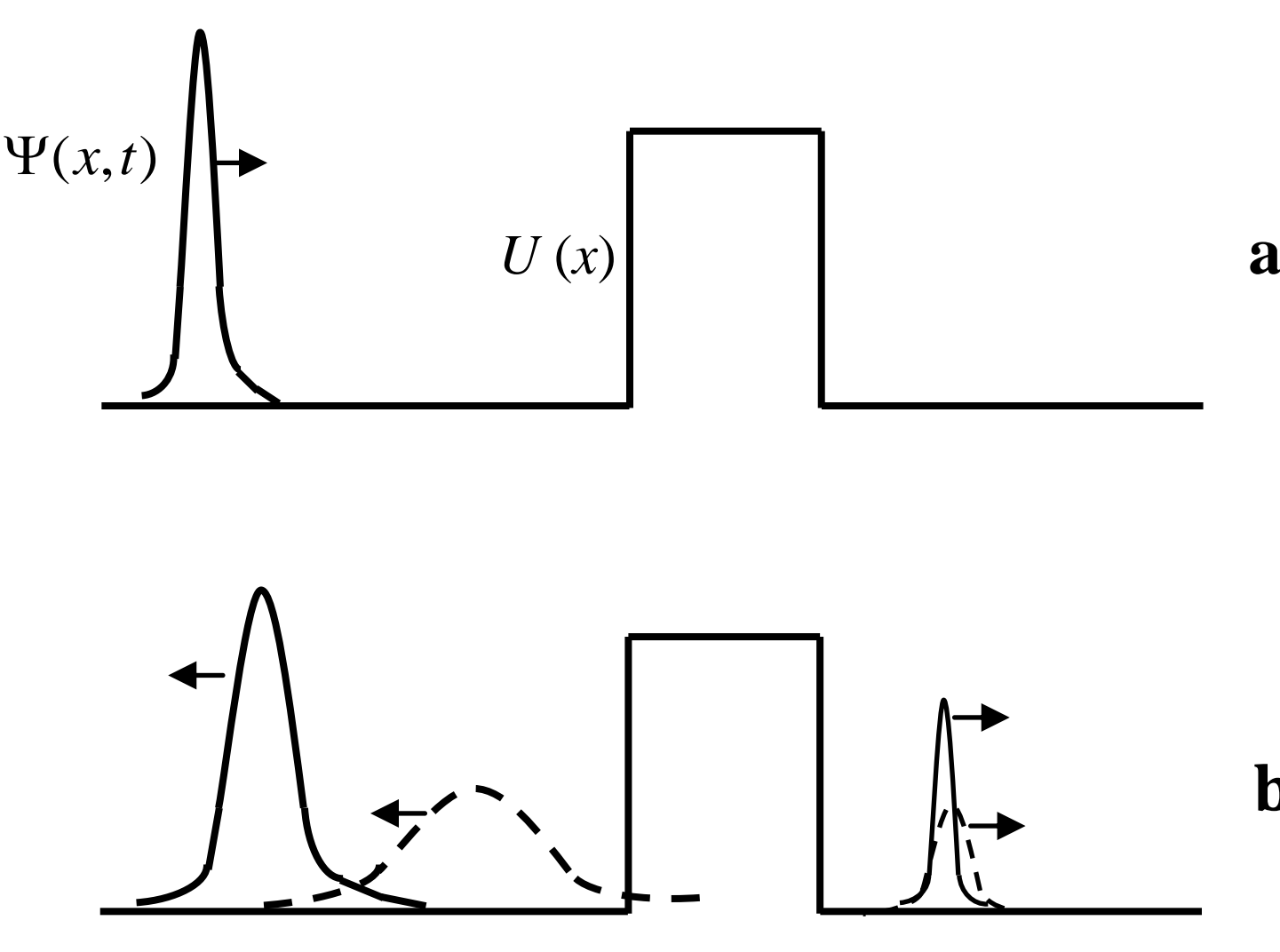

**Fig. 1**

Process of quantum tunneling with shown probability curves (not to scale)

**(a)** Input (a narrow incident packet);

**(b)** The first stage of the output. On the incidence side – first stage of reflection: combination of two waves reflected respectively from the front (solid curve) and back side of the barrier (dashed curve). The net first-stage reflection will be the sum of the *respective amplitudes*.

On the transmission side – first stage of transmission: solid curve – the forward-scattered wave from the front side; dashed curve – the forward-scattered wave from the back side. The superposition of the respective amplitudes will give the net first-stage transmission.

The next consecutive stages in (b) are decreasing reiterations of the first stage. The observed net outcome (total reflected and total transmitted pulse) is the superposition of all respective terms.

If we send only one particle at a time, only one detector will fire in each trial. If this is the one on the incidence side, the particle is reflected (back-scattered), but we cannot say confidently whether it is due to scattering at the front or at the rear of the barrier. And if the detector on transmission side fired (forward-scattering), one cannot say in principle at which side of the barrier the detected particle was scattered. *We cannot separate even approximately the "entrance scattering" from*

*"exit scattering" in the transmitted pulse.* And in many realistic cases when the incident packet width is comparable with the barrier width, we cannot do it even with the reflected pulse, either.

Mathematically, the role of scattering at the exit is seen from a very simple observation. Without the back side, the barrier becomes a potential threshold of height $U_0$, and we have for each frequency $\omega$ such that $\hbar\omega < U_0$ only one evanescent wave

$$\tilde{\Psi}(x,\, t) = \tilde{\mathcal{F}}_1(\omega)\, e^{-\kappa x} e^{-i\omega t}, \qquad x \ge 0 \tag{2}$$

with real positive $\kappa$. The amplitude $\tilde{\mathcal{F}}_1$ is determined from the boundary conditions at the threshold. There is no flux along the $x$-direction in this case. Putting (2) into the known expression for the flux density

$$j = \frac{\hbar}{\mu} \mathrm{Im}\left( \Psi \frac{\partial \Psi^*}{\partial x} \right) \tag{3}$$

gives

$$j = \frac{\hbar}{\mu} \kappa\, \mathrm{Im}\,(\tilde{\mathcal{F}}_1 \tilde{\mathcal{F}}_1^*)\, e^{-2\kappa x} = 0 \tag{4}$$

Here $\mu = E/c^2 = \hbar\omega/c^2$ is relativistic mass of the given particle. The result (4) is self-evident, since for the considered frequencies we have total reflection, with the zero net flux. But if we have the barrier of width *d* instead of a threshold, its back side automatically comes into play and brings in the second, exponentially increasing solution of the corresponding wave equation within the barrier, so instead of (2) we will now have

$$\Psi(x,t) = \left[ \tilde{\mathcal{F}}_1\, e^{-\kappa x} + \tilde{\mathcal{F}}_2\, e^{\kappa x} \right] e^{-i\omega t}, \qquad 0 \le x \le d \tag{5}$$

The corresponding amplitude $\tilde{\mathcal{F}}_2$ is determined by the boundary conditions at the exit.
In this case expression (3) gives

$$j = \frac{\hbar}{\mu} \kappa\, \mathrm{Jm}\,(\tilde{\mathcal{F}}_1 \tilde{\mathcal{F}}_2^* - \tilde{\mathcal{F}}_2 \tilde{\mathcal{F}}_1^*) = 2 \frac{\hbar}{\mu} \kappa\, \mathrm{Jm}\,(\tilde{\mathcal{F}}_1 \tilde{\mathcal{F}}_2^*) \ne 0 \tag{6}$$

Amplitude $\tilde{\mathcal{F}}_2$ represents scattering at the back side. It is this scattering that produces the second term in (5) and leads to a non-zero flux (tunneling!) through the barrier. There would be no tunneling without it!

All this makes equating the tunneling time to "*scattering time at the barrier entrance*" totally unsubstantiated.

**2.2 *Tunneling time and the wave frequency***

Let us now turn to a totally different expression for TT used in the same work [1]. In the "Discussion" section, the authors first mention the "…*empirical universal relation*" $\tau \approx (1/\nu) = T$ where $\nu$ is the wave frequency and *T* its period [4, 5], and then substitute it with the "…*general analytical expression*" (Eq. (4) in [1])

$$\tau = \frac{1}{\nu} \cdot A\,, \qquad \text{(A)}$$

which was introduced by S. Esposito [29]. Here *A* is a factor "…*depending on the physical barrier in question*". The authors then give the equation (Eq. (5) in [1]) for determining *A*:

$$\tau = \frac{\hbar}{\sqrt{E\,(U_0 - E)}} = \frac{1}{\nu}\,\frac{E}{4\pi^2 (U_0 - E)} \qquad \text{(B)}$$

where $E = \hbar\omega = 2\pi\hbar\nu$ is the particle's energy and $U_0$ is the height of the barrier (I use here notation $U_0$ instead of $V_0$ in the original text). Then combining (A) and (B) gives

$$A \equiv \frac{E}{4\pi^2 (U_0 - E)} \qquad \text{(C)}$$

But it is immediately seen that (B) is mathematically wrong, and it cannot give a consistent description of any physical situation. In particular, at $E \to U_0$ it yields $\tau \to \infty$ in blatant contradiction with the authors' own claim about universal finite (and very small!) tunneling time. At $E > U_0$ (passing above the barrier) the right-hand side of (B) gives negative passing time, whereas the middle part gives the imaginary time!

The only possibility to save (B) is to assume that it is an equation for *E* working only within the range $0 < E < U_0$ and selecting some specific value $E = E_s$ as its solution under certain conditions. But no such conditions are mentioned by the authors. Quite the contrary, the whole discussion claiming universal generality of its results implies that *E* is an independent variable that can be chosen arbitrarily in various possible experimental setups.

Even if we just accept (B) as a rule that selects some "legal" value $E = E_s$, the result will contradict some other statements in [1]. Indeed, a straightforward solution of (B) gives

$$E_s = \frac{4\pi^2}{1 + 4\pi^2}\, U_0 \qquad \text{(D)}$$

Putting this into (C) yields $A = A_s \equiv 1$. How can the unity factor be "..*depending on the physical barrier in question*"?

Next, according to (A), the "*universal quantum tunneling time*" must be given by

$$\tau_s = A_s\, \nu^{-1} = T \qquad \text{(A*)}$$

In a specific experiment described in [1], the input wave has $T = 115\,ps$, and the measured *experimental* value is $\tau = 130\,ps$. This is indeed, close to (A*), but only for specially selected energy value (D). Was (D) actually satisfied in the experiment? And what about other energies?

In the recent work [3], in the end of the Background section, the authors use the term "*universal interaction time*" instead of TT, and take $A = 1$ without any explanations.

Summarizing this part, we see that the two definitions of TT given in [1] are incompatible, indicating inconsistency in presented theoretical description. Even if we agree to consider the scattering only on the front of the barrier, its duration must be close to the time it takes the incident packet to pass through the "front door". This time is usually much longer than the period $T$ of the central monochromatic component of the packet. And it is longer still if we take into account multiple scattering. So the two definitions give widely different values for TT.

**2.3 *Tunneling in frustrated internal reflection***

In the description of experiment with the "…*symmetrical beam design of double prisms*" (Fig. 2) it is stated that "…*Reflected and transmitted signals were detected at the same time in spite of the fact that the transmitted beam traveled an additional distance d due to the gap*."

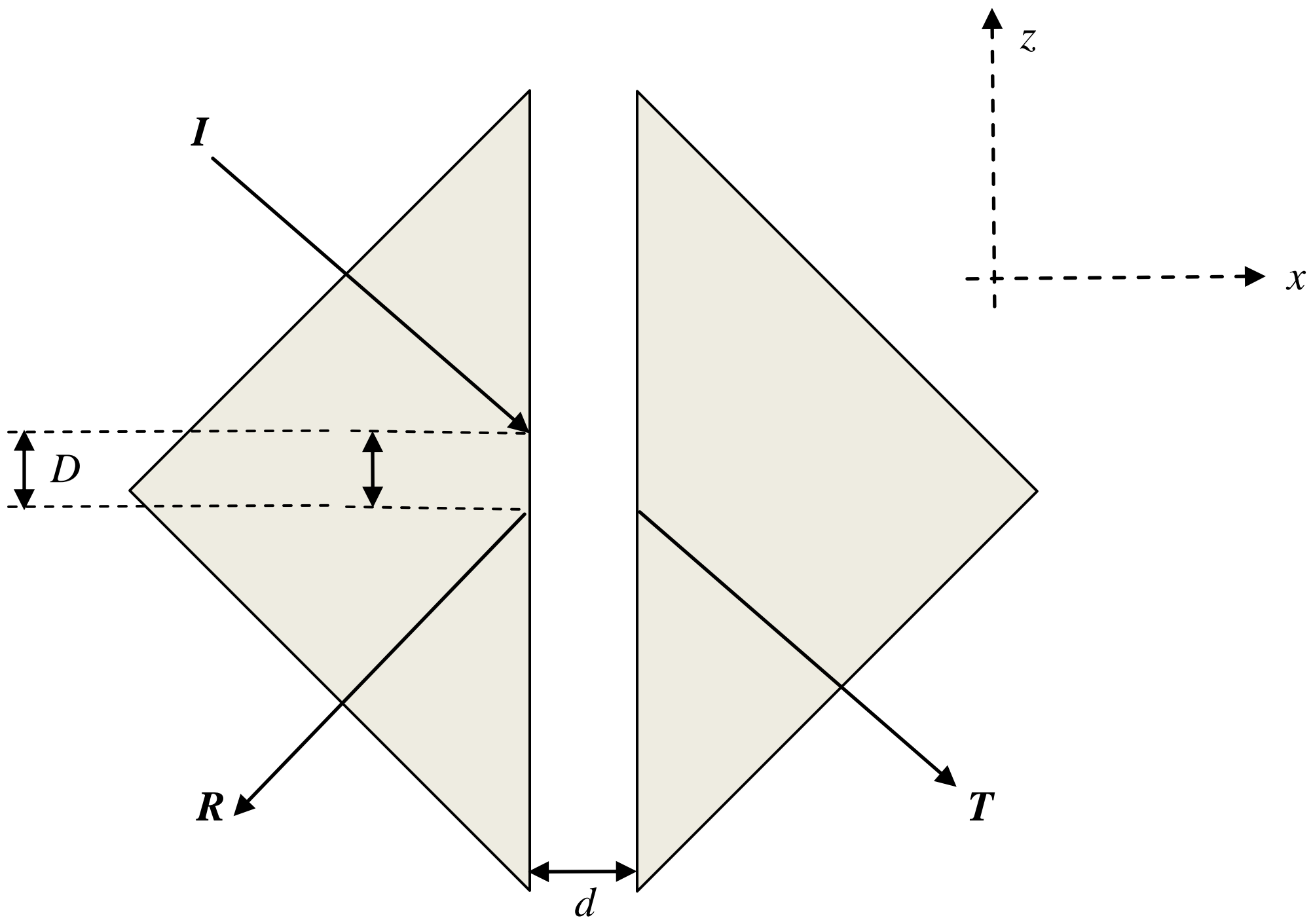

**Fig. 2** Sketch of the SQT experimental setup (following [1]).
Actually, it is known schematic of a beam-splitter. The gap between the prisms acts as a potential barrier of width *d*. ***I***, ***R***, and ***T*** are, respectively, incident, reflected and transmitted light pulses. ***R***, apart from reflection, undergoes the Goos-Hanchen shift by distance $D \sim \kappa_x^{-1}$, where $\kappa_x$ is the imaginary component of the wave vector **k** inside the barrier. Actually, ***R*** and ***T*** are formed on both sides of the barrier in the process of multiple scattering as described in Sec. 2.1.

The above-cited statement is based on two implicit assumptions: 1) the reflected part of the beam was only sliding by distance $D$ (the Goos-Hanchen shift) along the front of the gap without venturing inside, and 2) the transmitted part first undergoes the same shift together with the reflected part, and only after that turns by 90° to travel additional distance *d* across the gap in Fig. 2. Such artificial interpretation has nothing to do with real physics. A wave does not travel as a classical particle along a definite, specially selected path. A quantum-mechanical (QM) particle possesses wave properties and accordingly takes simultaneously all possible virtual paths between two locations of interest [24, 28, 30]. This is consistent with the reflection-transmission process described above in Sec. 2.1, according to which *each* of the two output beams in Fig. 2 is formed in the process of multiple scattering on both sides of the gap. Under the experimental conditions shown in Fig. 2 (the shift *D* and the gap *d* are of the same order of magnitude), both sides of the gap contribute approximately equally to the reflected and transmitted amplitudes. This makes the travel distances of both outputs practically equal in the described setup. And the same is true regarding the travel times – precisely what was observed in the experiment. The comparison of the two outputs under given conditions does not measure the TT, and in no way can its results be interpreted as the instantaneous tunneling, let alone instantaneous signaling.

### 2.4 *The frequency band of a signal*

One of the basic statements in [1, 3] is: "*…physical signals are frequency band limited*". The same is claimed by the authors nearly in all their publications on SQT. The argument used to support this claim is: "*An unlimited frequency band demands an infinite energy, since a field does only exist if it has at least one quantum at the frequency in question*" [3].

The concept of a signal is so important that these assertions require exhaustive scrutiny. Do the authors really think that each frequency in a band is actually represented by a separate photon with this frequency? If so, an unlimited frequency band would indeed imply an infinite number of photons and thus an infinite energy. But the same would then hold for band-limited signals as well! Any real band, limited or not, is continuous and thus contains not just infinite, but innumerable (uncountable) set of various frequencies. Such a set is infinitely more powerful than the set of all integers [31–33]. And according to the authors' logics, this would require an innumerable set of photons, each with sharply defined frequency, even for a limited band.

Such a nightmare has nothing to do with reality. First, no frequency in a band requires a separate particle (e.g., a photon) with this frequency. The whole band may be embraced by a single particle. It is only the corresponding *probability density*, not the separate particle, which is assigned to each frequency. A non-zero probability density for a frequency $\omega = \omega'$ (or for energy $E = E' = \hbar\omega'$) does not mean an actual pre-existing quantum with this energy. Second, a physical state of any real particle has *two* different characteristics: the whole frequency band $\Delta\Omega$ spanned by its wave packet, and some non-zero indeterminacy (standard deviation) $\Delta\omega$. They are sharply distinct from one another. The general mathematical rule for standard deviation $\Delta\xi$ of any variable $\xi$ ranging through a domain $\Delta\Xi$ is

$$\Delta\xi < \Delta\Xi \tag{7}$$

For frequency, this means $\Delta\omega < \Delta\Omega$, and in most cases $\Delta\omega$ is finite even when $\Delta\Omega \to \infty$. In some models we do have both $\Delta\Omega \to \infty$ and $\Delta\omega \to \infty$, e.g., in a spatially localized state $\Psi(x) = \Psi_0\delta(x - x')$ (Dirac's $\delta$-function) which is an equally-weighted superposition of de Broglie waves with all possible frequencies. The particle's energy in such state is indeed, infinite, but only because all its virtual frequencies, no matter how high, have equal probability to actualize. But such a state is only a highly idealized (albeit mathematically useful) model of some real situations. In most cases, high frequencies have sufficiently low probabilities, so $\Delta\omega$ and the average energy $\bar{E} = \hbar\bar{\omega}$ remain finite. A good example is a normalized Gaussian-type photon state $\Phi(\omega) = \Phi_0 \exp\left(-\frac{(\omega-\omega_0)^2}{2(\Delta\omega)^2}\right)$, which has an infinite range $\Delta\Omega$ around central frequency $\omega_0$, and yet finite $\Delta\omega$ and finite $\bar{E} = \hbar\bar{\omega} = \hbar\omega_0$.

Another well-studied case describes quasi-stationary states (or the Gamow states), as well as the energy (frequency) distribution in radiation emitted (absorbed) by a system in a quantum transition [34-36]. The emitted (absorbed) signal may be a single photon, and the corresponding distribution (probability per unit frequency) is

$$\frac{d\mathcal{P}(\omega)}{d\omega} = \frac{1}{2\pi}\frac{\gamma_0}{(\omega-\omega_0)^2 + (1/4)\gamma_0^2} \tag{8}$$

Here $\omega_0$ is the resonance frequency and $\gamma_0$ is the half-width of the packet, often used in Optics instead of $\Delta\omega$ (in most known states $\gamma_0 \approx (1/2)\Delta\omega \ll \omega_0$). But the distribution itself is infinitely spread, $\Delta\Omega \to \infty$. As in the previous example, the packet described by (8), while having unlimited frequency band, has finite (and very small in case of one photon) energy expectation value $\bar{E} = \hbar\omega_0$. A similar expression describes the scattering cross section as a function of $\omega$ near resonance [37, 38].

All this is a common knowledge apparently ignored by the authors. They also seem to ignore the well-known properties of Fourier transforms explaining the different shape of the same packet in different bases, the cases of its reshaping during propagation in configuration space, and the difference between the signal and group velocities [39–46].

The most important feature is that a truly band-limited wave packet with the zero Fourier amplitudes $\mathcal{F}(\omega)=0$ beyond some finite region $\Delta\Omega$ in the frequency space will be infinitely extended in configuration space. *Such a packet cannot be used for unambiguous signaling* which requires a wave form with a sharp edge. A spatially localized wave packet with a sharp edge and zero intensity beyond a finite region of space will have an infinite momentum range and thereby an infinite frequency band [16, 39]. This enables it for signaling but does not imply an infinite energy. As noted above, in most cases we deal with particles having each a finite energy despite the fact that $\Delta\Omega \to \infty$. In all known physical situations the existing high-frequency spectral components decrease sufficiently fast to make $\Delta\omega$ and $\bar{E}$ finite.

As an illustration of behavior of forms with an edge, consider a typical process including the preparation (initial stage) and then release and propagation of a possible signal carrier. Let the initial state $(t=0)$ be:

$$\Psi(x,0) = \sqrt{\frac{2}{a}}\begin{cases} \cos k_a x, & |x| \le a/2, \quad k_a \equiv \pi/a \\ 0, & |x| > a/2 \end{cases} \tag{9}$$

This is a textbook case describing a particle in its ground state in a potential box. Its physical realization may be the electric footprint of a photon in the lowest mode in a Fabri-Perrot resonator of length *a*. Here $|x|$ is the distance from resonator's center along its optical axis. The corresponding *x*-indeterminacy (standard deviation $\Delta x$) is ([28], Sec. 13.2)

$$\Delta x = a\sqrt{\frac{1}{12}\left(1-\frac{6}{\pi^2}\right)} < a \qquad (10)$$

In accordance with general rule (7), this is less than the range *a* allowed for the particle's position. The same property can be proved for the momentum space as well. The Fourier transform of (9) is

$$\mathcal{F}(k_x) = \frac{1}{\sqrt{2\pi}}\int \Psi(x) e^{-ik_x x} dx = 2\sqrt{\pi a}\ \frac{\cos\frac{1}{2}ak_x}{\pi^2 - a^2 k_x^2} \qquad (11)$$

The corresponding spectrum

$$\left|\mathcal{F}(k_x)\right|^2 = 4\pi a\ \frac{\cos^2\frac{1}{2}ak_x}{\left(\pi^2 - a^2 k_x^2\right)^2} \qquad (12)$$

spans all $k_x$-space:

$$k_x \in \left(-\infty, +\infty\right) \qquad (13)$$

But the standard deviation $\Delta k_x$ remains finite. This becomes immediately evident if we take into account the corresponding probabilities. The probability (12) for state (9) to collapse to a stationary $k_x$-th mode rapidly falls off with increase of $\left|k_x\right|$. The probability to find *any* $k_x$ higher than some value $k_x' \gg k_a$ is

$$\tilde{\mathcal{P}}(k_x') = 2\int_{k_x'}^{\infty} \left|\mathcal{F}(k_x)\right|^2 dk_x \underset{k_x' \gg k_a}{\approx} \frac{8}{3}\frac{\pi}{\left(ak_x'\right)^3} \qquad (14)$$

It becomes negligible at sufficiently high $k_x'$, so the remote regions of $k_x$-space give only a small contribution to the average $\overline{k_x^2}$. The exact value $\overline{k_x^2}$ is easier to calculate in the *x*-representation:

$$\overline{k_x^2} = \int k_x^2 \left|\mathcal{F}(k_x)\right|^2 dk_x = \hbar^{-2}\int \Psi^*(x)\hat{p}_x^2\Psi(x)dx = -\frac{2}{a}\int_{-a/2}^{a/2} \cos k_a x \frac{\partial^2}{\partial x^2}\cos k_a x\, dx = k_a^2 \qquad (15)$$

Since (9) is an even function, we have $\overline{k}_x = 0$, so that

$$\Delta p_x = \hbar \Delta k_x = \hbar\sqrt{\overline{k_x^2} - \overline{k}_x^{\,2}} = \hbar k_a = \pi\frac{\hbar}{a} \qquad (16)$$

Thus, despite the infinite range (13) of $k_x$, the standard deviation and thereby QM indeterminacy $\Delta k_x$ is finite.

The values (15) and (16) turn out to be the same as the corresponding values for an unbound monochromatic standing wave with $\lambda = 2a$. This leads sometimes to the wrong statements that the mode (9) *is* represented by such a wave. But, in contrast with running or standing de Broglie's wave which occupies all space, state (9) is non-zero only within one restricted domain. And conversely, while a standing wave has only two sharply-defined non-zero momenta $k_x = \pm\, k_a$, Eq-s (11, 12) describe a continuous distribution whose spectrum only peaks at $k_x = \pm\, k_a$.

For a particle in a box, an interesting feature of such state is that despite of its indeterminate momentum (and thereby indeterminate kinetic energy!) with infinitely spread spectrum, it has definite *net* energy (frequency), simply because it is an eigenstate of the corresponding Hamiltonian. In case of a single photon inside the resonator, the corresponding frequency is

$$\omega = \frac{E}{\hbar} = \omega_a \equiv ck_a \qquad (17)$$

This is consistent with indeterminate momentum and kinetic energy $K$, since $E$ is the sum

$$E = K + U\,, \qquad (18)$$

in which the potential energy $U$ is also indeterminate, and both indeterminacies are correlated so that the sum (18) remains fixed. This fixed value is equal to the sum of the respective *expectation values* $\bar{K}$ and $\bar{U}$ which can be calculated for the given bound state. A momentum measurement will destroy state (9) since definite momentum is characteristic of a free particle. This state will be destroyed by position measurement as well, since a definite (even if only fleeting) position requires an agent (some external time-varying force) provoking the particle to collapse to this position.

Now we go to the second stage: we release the particle, for instance, by making the trapping box (resonator) transparent. The moment we do it, the energy becomes also indeterminate. The release operation *changes* the Hamiltonian, so the state is no longer stationary, and accordingly, the energy acquires a certain spread. Formally, this follows directly from (18): elimination of $U$ from the equation leaves indeterminacy $\Delta K$ in $K$ unbalanced and thus transforms it directly to $E$. After that, it is its expectation value $\bar{E}$ (together with standard deviation $\Delta E$) rather than energy as such that becomes characteristic of the resulting state. The value $\bar{E}$ remains finite and close to the initial eigenvalue (17); the energy indeterminacy, in case of a photon, trivially obtains from (16) as

$$\Delta E \equiv \hbar\Delta\omega = \hbar c \Delta k_x = \pi\frac{\hbar c}{a} = \hbar\omega_a \qquad (19)$$

The energy spectrum can be obtained from (12) by replacing $k_x \to \omega / c$. Again, $\bar{E}$ and $\Delta E$ are both finite while the frequency band, according to (11), (12), is now infinite.

All these properties are obtained under the assumption that the release operation, while changing the Hamiltonian, does not change the shape of the initial form. This condition can be met with high accuracy. But the form now acquires the ability to propagate beyond its initial confinement and thus can be used as a signal. Its farther evolution cannot be described by evolution operator which treats only continuous wave forms. But it can be found by using the new technique [28], which can treat even the forms with the most outrageous discontinuities, such as $\delta$-functions.

Summarizing this part, we see that contrary to the statements in [1, 3], the energy of a signal is finite in all practical cases, and the signal may pass through a barrier in a tunneling mode under condition $\bar{E} < U_0$. In order to see whether this process leads to SS, we can compare the tunneled packet with an identically prepared packet arriving at the same location along the path of equal length but without an intervening barrier. If in a considerable fraction of all trials the arrival time of the first signal is earlier than the *earliest* arrival of the second one, we could talk about superluminal signal transfer. According to analysis of known experiments [43, 44], such a result has not been shown.

**2.5 *The tunneling intensity***

Expression for the tunneling intensity given by Eq. (2) in [1]

$$I_t(x) = I_0 e^{(i\omega t - 2\kappa x)} \qquad \text{(E)}$$

is obviously wrong. The intensity is a directly measurable physical characteristic strictly proportional to probability. It cannot be a varying function of time for a state with definite $\omega$, let alone a complex function.

Recently the following objection was raised against this argument: "*It is not unusual to use complex equations for waves even if only the real part has a physical meaning*".[1]

This statement, using Pauli's famous expression, "…is not even wrong". It is much worse than just wrong, because its first part may sometimes be correct – but only for those characteristics that can be described by a complex expression. We know about complex permittivity, complex energy, complex refraction index and wave number, complex impedance, and so on. Moreover, we know that both - real *and* imaginary part of all these characteristics have physical meaning. We routinely use complex expressions for waves, and in QM, the wave function $\Psi(x,t)$ *must* be complex, see, e.g. [28]. But what is possible and sometimes even necessary for a wave function, is absolutely forbidden for the *wave intensity* which determines probability and/or concentration of particles appearing as one number in a single measurement. It is immediately seen that applying the presented objection to rescue (E) would be fatal. Taking the real part of (E) (assuming $I_0$ real) would yield

$$\operatorname{Re} I_t(x) = I_0\, e^{-2\kappa x} \cos\omega t \qquad \text{(F)}$$

Inadequacy of this expression is evident for any Physics junior. It has nothing to do with *intensity* of a monochromatic wave, which is spatially uniform, time-independent, cannot oscillate, and cannot take on negative values.

**2.6 *Energy-momentum relation***

The concluding part of [1] states that in the tunneling effect we observe "*The violation of the Einstein energy relation by the imaginary wave number* $E^2 = (\hbar k\, c)^2 + (m_0 c^2)^2$". This is a totally misleading statement because the above relation describes a free particle whose wave number cannot be imaginary. A tunneling particle is not free, and the energy relation for it is $\left(E - U_0\right)^2 = (\hbar k\, c)^2 + (m_0 c^2)^2$. The imaginary wave number $k = i\kappa$ emerges as its solution when $E < m_0 c^2 + U_0$, which is just the relativistic tunneling condition. The energy relation would also hold for free tachyons [6, 9-16], in which case it requires the imaginary rest mass, $m_0 = i|m_0|$. In neither case is the energy relation violated.

---

[1] An anonymous reviewer at the same journal that published [1] and [4]

### 3. Tunneling velocity and reshaping of tunneling wave packet

Here we discuss two definitions of quantum tunneling velocity (QTV) reflecting two different characteristics of the tunneling rate [39-44]. Both definitions refer to the shape of tunneling pulse.

In the first definition we prepare a packet with a sharp edge (e.g., by releasing a photon from the Fabri-Perrot resonator) and determine the edge's velocity. According to the Brillouin-Zommerfeld theorem [39], the edge is necessary for unambiguous signaling, but its motion can be only luminal or subluminal, $v_{edge} \le c$.

The second definition associates QTV with the velocity of maximum of the wave packet. It is close to its group velocity [21, 22] and can move FTL. One known mechanism for FTL motion is a specific reshaping of the packet in a non-equilibrium medium (some features of this process are described qualitatively in [6]). Due to reshaping in the tunneling process, the evolving top of the packet may get closer to its leading front, thus effectively exceeding the speed *c*. But this kind of superluminal motion cannot be used for signaling [39, 42-44].

Thus, the motion of some parts of the tunneling packet can be superluminal, but cannot be harnessed to transfer a signal. For the latter we need a sharp edge, but its motion cannot be superluminal. Nature seems to conspire against SS to maintain causality.

The authors of [1-5] deny the existence of an edge in a wave form and claim that in their experiments the tunneling pulse achieves superluminal velocity without reshaping. Even if that were 100% true, it would still fall short of SS, once there is no front with discontinuity. But the claim about absence of reshaping cannot be true. Except for only one case (light pulse in vacuum), reshaping is a universal effect. It consists of three contributions. The first one comes from the difference in optical characteristics of the mediums within and outside the barrier. It can be seen already on qualitative level when we consider, e.g., a photon entering the barrier. Let each medium be homogeneous and described by refraction index $n(\omega)$ outside and $\tilde{n}(\omega)$ within the barrier. At the interface, all monochromatic components of photon's wave packet undergo change of their respective wavelengths $\left(\lambda \to \tilde{\lambda}\right)$ due to the change $n \to \tilde{n}$. This results in the corresponding change (spread or contraction) of the packet's shape after its passing through the interface. Apart from this obvious effect, there is the second contribution due to the change of the components' *amplitudes* in the reflection–transmission process at the boundary: $\mathcal{F}(k_x) \to \tilde{\mathcal{F}}(\tilde{k}_x)$. The process is $k_x$-dependent, so each $\mathcal{F}(k_x)$ undergoes its own individual change. This changes spectral composition of the transmitted (and reflected!) fractions of the initial packet, and thus their respective shapes. One cannot dismiss these two types of reshaping as merely a boundary effect, because a barrier may itself contain many boundaries inside when it is formed by a multi-layered periodic structure.

The third contribution, most relevant to the discussed phenomenon, is due to dispersion even inside of a *homogeneous* barrier. The dispersion may be small for a frequency range far from the absorption bands, but it is never zero. And in the quantum tunneling experiments, dispersion is usually a dominating factor, even if tunneling takes place through a vacuum barrier. In particular, for tunneling in frustrated total reflection experiment as shown in Fig. 2, the pulse within the gap can be represented as

$$\tilde{\Psi}(x,t)=\int\left[\tilde{\mathcal{F}}_1(\tilde{k}_x)e^{i\tilde{k}_x x}+\tilde{\mathcal{F}}_2(\tilde{k}_x)e^{-i\tilde{k}_x x}\right]e^{-i\omega t}d\tilde{k}_x \qquad (20)$$

Here

$$\tilde{k}_x=\sqrt{\frac{\omega^2}{c^2}-\tilde{k}_z^2}\ , \qquad (21)$$

and $\tilde{k}_z$ is determined from the boundary condition on the prism-gap interface with the angle of incidence $\theta=\pi/4$,

$$\tilde{k}_z=k_z=\frac{\omega}{c}n\sin\theta=\frac{1}{\sqrt{2}}\frac{\omega}{c}n \qquad (22)$$

Putting this into (21) gives

$$\tilde{k}_x=\frac{\omega}{c}\sqrt{1-\frac{1}{2}n^2} \qquad (23)$$

For a typical refraction index $n\approx 1.5$ of a prism we have

$$\tilde{k}_x=i\alpha\frac{\omega}{c} \quad \text{with} \quad \alpha\equiv\sqrt{\frac{1}{2}n^2-1}\ \approx 0.35 \qquad (24)$$

Thus, (20) can be written in terms of frequency as

$$\begin{aligned}\Psi(x,t)&=i\frac{\alpha}{c}\int\left[\tilde{\mathcal{F}}_1\left(i\alpha\frac{\omega}{c}\right)e^{-\alpha\frac{\omega}{c}\left(x+i\frac{c}{\alpha}t\right)}+\tilde{\mathcal{F}}_2\left(i\alpha\frac{\omega}{c}\right)e^{\alpha\frac{\omega}{c}\left(x-i\frac{c}{\alpha}t\right)}\right]d\omega=\\&=\tilde{\Phi}_1\left(x+i\frac{c}{\alpha}t\right)+\tilde{\Phi}_2\left(x-i\frac{c}{\alpha}t\right)\end{aligned} \qquad (25)$$

This is totally different from a textbook case $\Psi(x,t)=\tilde{\Phi}\left(x\pm ct\right)$ for a fixed-shape pulse propagation in a free space (the only reshaping-free exception mentioned above). In contrast to such case, here we have superposition of two packets "propagating" in the opposite directions with superluminal but imaginary speed $i\tilde{u}\equiv ic/\alpha$. The actual propagation without quotation marks appears as the result of superposition of the *two* opposite and phase-shifted evanescent waves. The resulting single tunneling pulse may have superluminal group velocity and will definitely change its shape. The latter becomes obvious if we rewrite the right-hand side of (25) as

$$\begin{aligned}&\tilde{\Phi}_1\left(x+i\frac{c}{\alpha}t\right)+\tilde{\Phi}_2\left(x-i\frac{c}{\alpha}t\right)=\\&=\tilde{\Phi}_1\left(x+ut+\left(i\tilde{u}-u\right)t\right)+\tilde{\Phi}_2\left(x-ut-\left(i\tilde{u}-u\right)t\right)\end{aligned} \qquad (26)$$

Due to additional complex terms in the arguments of $\tilde{\Phi}_1\left(x+i\dfrac{c}{\alpha}t\right)$ and $\tilde{\Phi}_2\left(x-i\dfrac{c}{\alpha}t\right)$, the corresponding wave forms vary with time, and so does their sum describing the whole packet.

Note that this result obtains even under the simplifying assumption that the medium is non-dispersive ($n$ and thereby $\alpha$ are approximated by constant numbers independent of $\omega$). Taking account of dispersion will make the reshaping even more pronounced.

A small reshaping exists already in regular propagation of light in a transparent medium like glass. The result (26) shows how tunneling gives rise to reshaping even in vacuum, which is the single non-dispersive medium, and even that – only for photons. All the massive particles undergo dispersion in vacuum as well, so that their wave forms evolve even at regular propagation [28, 45, 46], let alone at tunneling. Claims about quantum tunneling without reshaping are mathematically and physically incorrect.

## 4. SQT and causality

Finally, we consider the relation between SQT and causality as presented in [1-5], and discuss a thought experiment which is a specific version of the Tolman paradox [6, 15-17], showing unambiguous connection between SS and causality violation.

If the tunneling is superluminal but cannot be used for signaling, it will just add a new case to a compendium of known FTL motions. Interesting as it may be, it will not change our vision of the world. But if it turns out to be suitable for SS, then it would probably be the greatest discovery in history, shattering the foundations of all Physics. The realization of SS would mean the end of causality as a universal principle, and could outline some essential restrictions on it.

The most general conventional definition of causality is "*The cause precedes the effect*." Its equivalent formulation in terms of SR is: "*No signal can propagate faster than the speed of light in vacuum*."

The authors of [1-5] deny the equivalence of these two definitions. They call first of them "*the primitive causality*", and the second one – "*the Einstein or relativistic causality*". The claim is that SS "*violates only the Einstein causality, but not primitive causality*." This implies that the Einstein causality is nothing more than just the ban on SS, and "lifting" this ban would have no consequences in anything else. Obviously, the authors ignore the relativity of simultaneity and its unavoidable implications (the Tolman paradox) that would arise from SS.

Let us consider these consequences, assuming the reality of SS but keeping in mind that such an assumption is so far a pure speculation. So accept for a while two mutually excluding assumptions: the tunneling pulse has a sharp edge, *and* its velocity $v_{edge} \equiv \tilde{v} > c$. *Then* the pulse can be used for SS. Assume also that the effects of attenuation of the pulse within the barrier can be neglected. Denote the pulse entrance into the barrier as event A, and its exit at the back side as event B. Since the pulse carries a signal, A and B are causally connected: e.g., B may be an encoded text of Moonlight Sonata received only due to its sending from A. And due to superluminality of the signal carrier, the pair of events (A, B) forms a space-like 4-interval. This fact and the corresponding violation of causality could, in principle, be observed in a 3-stepped extension of the discussed experiments. The first step is designed to record the time ordering of (A, B) in two different inertial reference frames: one is the Lab, with observer Lucy, and the other, denoted as S, with observer Sam. S must move relative to the Lab at a speed $V > c^2/\tilde{v}$ in the same direction as the tunneling pulse, and must be receding from L by the beginning of the experiment. Under these conditions, the pair (A, B) would be observed in S in the reversed order – as (B, A): the text of Moonlight Sonata spontaneously emerges at the exit of the receding barrier, tunnels through it faster than light, and reaches the front, where it merges with its original "self" just arriving from the Lab source to enter the barrier. This already looks like a miracle to Sam (see more details in [6, 9-14]). And add to it the fact that the concentration of tunneling bits as measured by Sam rapidly increases from B to A! This would be really weird, but theoretically possible since a chance of such an event, albeit exponentially small, is still not zero. But if we repeat the experiment a few times and the described weird process reiterates at each trial, the evidence of SS becomes 100% compelling. In each trial the Moonlight Sonata tunnels through the barrier – and in S the signal first emerges spontaneously at the back of the barrier and then moves to its front.

Already this step shows intimate connection between SS and causality. Sam knows from previous communications with Lucy that A is the cause of B. Therefore his observing B preceding A is already violation of causality in its first (general) formulation. In S, the effect (B) precedes its cause (A). The situation is mind-boggling for Sam, but so far nothing unusual (except for SS itself) is observed by Lucy in her Lab.

Now add the second step: let Sam, apart from passive recording the described events, copy the Moonlight Sonata right after its spontaneous emission at the back of the L-barrier, and sends it to Lucy at a speed $\tilde{v}' \geq \tilde{v}$ through another barrier which is *stationary in* S. This time it will be Lucy's turn to get surprised. Under conditions of the Tolman paradox the copied musical message from Sam may reach her not only before Sam sends it, but even *before* she produces the initial message. This will be observed by her as spontaneous emission of the Moonlight Sonata before she has selected this specific piece for signaling, that is, before the beginning of the first step. As at the first step, attempts to explain this as an extra-rare but theoretically possible fluctuation will not work, because the same situation would reiterate in each independent trial. This will be already a full-fledge violation of causality in its first formulation, and for both observers.

Finally, we can add the third step: Let the premature spontaneous emission of the very same sonata that Lucy is preparing to send to Sam makes her to cancel the planned communication. Then we arrive at a real paradox: if Lucy does cancel it, then the whole chain disappears, there are no spontaneous emissions, so Lucy sends the signal. If she sends it, there follow spontaneous emissions, and she does not send it. In the framework of classical thinking, this is logical contradiction, showing that the initial assumption about possibility of SS was false.

A possible resolution in the framework of QM was suggested in [15]: allowing Lucy to be in an entangled superposition of action and inaction with her environment, mathematically identical to the similar state of the Schrodinger cat. But even if such superposition on the macroscopic scale could exist, so far nobody can tell what it would look like.

The situation becomes even more complicated if we take account of the so far neglected attenuation of the signal within the barrier. The signal becomes exponentially weaker with increase of the barrier size, and the latter must exceed a certain value for the time loop in the third step to become noticeable.

Under all these requirements, the described experiment can hardly be carried out with today's technology. But if performed someday *and* showing possibility of SS, it may open new dimensions in our understanding of causality and the world. And in any event, it shows that contrary to the statements in [1-5], SS would violate causality in all its current formulations.

## Conclusion

Most of the presented arguments about SQT in [1-5] are either incoherent or just wrong. The experiments on quantum tunneling described there are interesting in their own right and may lead to the new insights into the phenomenon of signal transfer. But as has been pointed out in [8] on another occasion, precisely for this reason, the obtained results and their interpretation need thorough scrutiny. As of now, the claims presented in [1-5] are not substantiated. We need more research, both – theoretical and experimental – to reach a better understanding of SQT.